\documentclass[12pt]{article}
\usepackage{amsmath}
\usepackage{amsfonts}
\usepackage{amssymb}
\usepackage{graphicx}

\def\ffrac#1#2{\textstyle\frac{#1}{#2}\displaystyle}

\begin{document}
\setcounter{page}{0} \topmargin 0pt
\renewcommand{\thefootnote}{\arabic{footnote}}
\newpage
\setcounter{page}{0}

\begin{titlepage}

\begin{center}
{\Large {\bf Some Results on Mutual Information of Disjoint Regions in Higher Dimensions}}\\

\vspace{2cm}
{\large John Cardy$^{a,b}$\\}  \vspace{0.5cm} {\em $^{a}$Rudolf
Peierls Centre for Theoretical Physics\\ 1 Keble Road, Oxford OX1
3NP, UK\\}  \vspace{0.2cm} {\em $^{b}$All Souls College, Oxford\\}

\vspace{2cm}

\end{center}

\vspace{1cm}

\begin{abstract}
\noindent We consider the mutual R\'enyi information $I^{(n)}(A,B)\equiv S^{(n)}_{A}+S^{(n)}_{B}-S^{(n)}_{A\cup B}$ of disjoint compact spatial regions $A$ and $B$ in the ground state of a $d$+1-dimensional conformal field theory (CFT), in the limit when the separation
$r$ between $A$ and $B$ is much greater than their sizes $R_{A,B}$. We show that in general $I^{(n)}(A,B)\sim C^{(n)}_AC^{(n)}_B(R_AR_B/r^2)^{\alpha}$, where $\alpha$ is the smallest sum of the scaling dimensions of operators whose product has the quantum numbers of the vacuum, and the constants $C^{(n)}_{A,B}$ depend only on the shape of the regions and universal data of the CFT. 

For a free massless scalar field, where $\alpha=d-1$, we show that $C^{(2)}_AR_A^{d-1}$ is proportional to the capacitance of a thin conducting slab in the shape of $A$ in $d$+1-dimensional electrostatics, and give explicit formulae for this when $A$ is the interior of a sphere $S^{d-1}$ or an ellipsoid. For spherical regions in $d=2$ and $3$ we obtain explicit results for $C^{(n)}$ for all $n$ and hence for the leading term in the mutual information by taking $n\to1$. We also compute a universal logarithmic correction to the area law for the R\'enyi entropies of a single spherical region for a scalar field theory with a small mass. 

\end{abstract}

\end{titlepage}

\section{Introduction}

Since the pioneering paper of Srednicki \cite{sred} there has been increasing interest in understanding and quantifying entanglement in quantum field theories. In that paper it was shown that, in a free scalar field theory, the von Neumann entropy $S_A=-{\rm Tr}\,\rho_A\log\rho_A$ of the reduced density matrix $\rho_A$ describing the degrees of freedom inside a spherical region $A$, which measures the entanglement of the degrees of freedom in $A$ with those in its complement, is proportional to the area of its boundary. Subsequently this `area law' was show to be generic in space dimensions $d\geq2$ \cite{arealaw}, and this prompted comparisons with black hole physics. 

However in 1993 Holzhey \em et al.~\em \cite{holzhey} showed that in a conformal field theory (CFT) in $d=1$ the entanglement entropy of an interval of length $R_A$ goes like $\log R_A$ with a universal coefficient proportional to the central charge $c$. (Similar logarithms are now understood to occur whenever $d+1$ is even, but for higher $d$ these are non-leading with respect to the area term \cite{logs}.) Subsequently this logarithmic behaviour was observed in numerical studies of critical quantum spin chains whose long-distance behaviour is believed to be described by a CFT \cite{spinchains}. A more complete analysis of entanglement in 1+1-dimensional CFTs was given in 
Refs.~\cite{cc}. 

More recently \cite{cct} these methods, which involve using the so-called replica trick of computing $S_A$ as limit as $n\to1$ of the R\'enyi entropies
$$
S_A=\lim_{n\to1}S_A^{(n)}\quad\mbox{where}\quad S_A^{(n)}=(1-n)^{-1}\log{\rm Tr}\,\rho_A^n
$$
have been extended to the computation of the entanglement entropy $S_{A\cup B}$ between two disjoint intervals $A,B$ and the rest of the system in a 1+1-dimensional CFT. It was shown that this encodes all the data of the CFT, not only the central charge. Moreover the mutual information, given by the limit as $n\to1$ of the difference of R\'enyi entropies
$$
I^{(n)}(A,B)\equiv S^{(n)}_{A}+S^{(n)}_{B}-S^{(n)}_{A\cup B}
$$
has an expansion of the form 
\begin{equation}\label{eq:1}
I^{(n)}(A,B)=\sum_{\{k_j\}}C^{(n)}_A(\{k_j\})C^{(n)}_B(\{k_j\})\left(\frac{R_AR_B}{r^2}\right)^{\sum_{j=1}^nx_{k_j}}\,.
\end{equation}
Here $R_A,R_B$ measure the lengths of the two intervals, and $r$ is the separation between their centres. The sum is over a set of scaling operators of the CFT, labelled by $k_j$, with scaling dimension $x_{k_j}$, for each replica $j$.
The coefficients are universal and encode information about the correlation functions of these operators in the plane. 
For $R_AR_B/r^2\ll1$ the leading term in (\ref{eq:1}) comes from the case when only two of the $x_k$ are non-vanishing and correspond to the lowest dimension operator in the CFT. It was possible \cite{cct} to continue this result analytically in $n$ to compute the leading term in the mutual information $I(A,B)$. 

The one-dimensional case has also been studied numerically in a number of papers \cite{1d}.

The arguments of Ref.~\cite{cct} were based on a kind of operator product expansion first used by Headrick \cite{head}.
In this paper we argue that (\ref{eq:1}) also holds for the mutual R\'enyi entropies in higher dimensional CFTs. However in this case the coefficients are much more difficult to compute, and we succeeded in obtaining explicit results for the leading term only in the simple case of a free massless scalar field theory, and when $A$ and $B$ are spheres of radii $R_A$ and $R_B$. In principle our methods work in any (integer) number of dimensions, but the results are most simply expressed in $d=2$ and 3.

In general we find
$$
I^{(n)}(A,B)\sim g^{(n)}_d\left(\frac{R_AR_B}{r^2}\right)^{d-1}\,,
$$
where, in 3+1 dimensions,
$$
g^{(n)}_3=\frac{n^4-1}{15n^3(n-1)}\,.
$$
For the mutual information ($n=1$) this gives $g^{(1)}_3=\frac4{15}=0.26\dot6$, to be compared with a numerical result  $0.26$ due to Shiba \cite{shiba}. 

In 2+1 dimensions  
\begin{eqnarray*}
g^{(n)}_2&=&\frac{n}{2\pi^2(n-1)}\int_0^\infty\int_0^\infty\bigg(\frac{(1-x)(1-y)(1-(xy)^{n-1})}{1-xy}\\
&&\quad\qquad\qquad\qquad\qquad\qquad+(1-x^{n-1})(1-y^{n-1})\bigg)\frac{dxdy}{(1+x)(1+y)(1-x^n)(1-y^n)}\,,
\end{eqnarray*}
which leads to $g^{(1)}_2=\frac13$, to be compared with the numerical result $0.37$ \cite{shiba}. 

We are also able to compute the form of the corrections to the leading term, which should be $O(R_AR_B/r^2)^{2(d-1)}$
and also $O(R_AR_B/r^2)^{d+1}$. In principle the coefficients are calculable. Note that these are more important in $d=2$ than in $d=3$ which may account for the above discrepancy. 

Our methods use the conformal invariance of the massless free field theory, which in fact implies that for $R_A\not=R_B$
the actual expansion parameter is
$$
\frac{R_AR_B}{r^2-(R_A-R_B)^2}\,.
$$
In this form, our results also apply to the mutual entanglement between the interior of a sphere of radius $R_A$ and the exterior of a concentric sphere of radius $R_B$, in the limit when $R_A\ll R_B$.

For $n=2$ and general shapes we show that the coefficients $C^{(2)}_{A,B}$ are given by the capacitance of $d$-dimensional bodies $A,B$ in $d+1$-dimensional electrostatics. 

The mutual information for a free scalar field in higher dimensions has been studied in only a few papers. Casini and Huerta \cite{ch} and Shiba \cite{shiba0} showed that it should decay as $(R_AR_B/r^2)^{d-1}$ at large separations (and  $(R_AR_B/r^2)^d$ for free fermions \cite{ch}).  This was based on the expression for the density matrix in terms of the correlation functions \cite{bomb} which holds for any system with a gaussian wave functional. However the coefficients must still be determined numerically for finite separations and then extrapolation. Recently this was carried out by Shiba \cite{shiba} for the case of equal spheres in $d=2$ and $3$ space dimensions, and also rings and shells  \cite{shiba} as well as some other shapes \cite{shiba2}. This shows that the mutual information is not extensive, that is, is not given by a double integral of a kernel over the boundaries or the bulk of $A$ and $B$. 

Free and interacting fermions at finite chemical potential (Fermi liquids) have been studied in \cite{swin}, and the entanglement of the radiation field with a dielectric medium in \cite{klich}.

The layout of this paper is as follows. In Sec.~2 we first recall the expressions for the various R\'enyi entropies $S^{(n)}_R$ as path integrals over $n$ copies of ${\mathbb R}^{d+1}$ sewn together in a particular way along the boundary of the selected region $R$ to form the conifolds ${\cal C}^{(n)}_R$. We then consider the general case of the small $R_{A,B}$ expansion, and show how the coefficients $C^{(n)}_{A,B}$ in this expansion are related to one- and two-point functions on ${\cal C}^{(n)}_{A,B}$.

The rest of the paper is devoted to the special case of a massless free scalar field theory. In Sec.~3.1 we show that for $n=2$ the coefficients $C^{(2)}_{A,B}$ are related to the electrostatic capacitance of each region. In an Appendix we derive explicit formulas for the case where $A$ and $B$ are spherical, or more generally ellipsoidal, by generalising a famous method due to W.~Thomson to general dimension $d$. In Sec.~3.2 we obtain results for the spherical case for general $n$ and $d$,
using conformal invariance. We show that for $d$ odd the coefficients are polynomials in $n$ which can then be continued to $n\to1$ to find the von Neumann entropy. For $d$ even the result is not a polynomial but explicit results may nevertheless be obtained for $n$ integer as well as the limit $n\to1$. In Sec.~3.3 we consider the case of a single spherical region in a free massive scalar theory, and confirm the existence of universal logarithmic terms in the R\'enyi entropies first predicted in \cite{cc}. 

\section{General form of the expansion for $R_{A,B}\ll  r$.}
Consider a $d$+1-dimensional quantum field theory in a domain $D$, in its ground state $|0\rangle$. In this paper we consider $D$ to be ${\mathbb R}^d$, although some of the results may be adapted to semi-infinite and finite domains and also to finite temperature. 

If $X$ is some subdomain, we suppose that, in the presence of a suitable UV cut-off, the Hilbert space can be decomposed as ${\cal H}_X\otimes{\cal H}_{\overline X}$. The R\'enyi entropies are $S^{(n)}_X=(1-n)^{-1}\log P^{(n)}_X$, where 
$$
P^{(n)}_X={\rm Tr}_{{\cal H}_X}\,\rho_X^n\qquad\mbox{with}\qquad\rho_X={\rm Tr}_{{\cal H}_{\overline X}}\,|0\rangle\langle0|\,.
$$

As explained in Ref.~\cite{cc}, these can be expressed in terms of a path integral on a particular conifold as follows. The ground state wave-functional $\langle\{\psi(x)\}|0\rangle$ is given by a path integral in imaginary time on the half-space ${\mathbb H}_-={\mathbb R}^d\times(-\infty,0)$ from $\tau=-\infty$ to $\tau=0$, with the fields constrained to take the values $\{\psi(x)\}$ on $\tau=0$. Similarly 
$\langle0|\{\psi(x)\}\rangle$ is given by the path integral on ${\mathbb H}_+$ from $\tau=0$ to $\tau=\infty$. Each of these should be normalised by $Z^{-1/2}$, where $Z$ is the partition function on ${\mathbb R}^{d+1}$. [Strictly speaking we should restrict the integration to $|\tau|<T$ before taking the ratio, then let $T\to\infty$, and similarly with the thermodynamic limit in space.]

To get $P^{(n)}_X$ we take $n$ copies of ${\mathbb H}^{(j)}_\pm$  labelled by $j=0,\ldots,n-1$, and sew ${\mathbb H}_-^{(j)}$ to 
${\mathbb H}_+^{(j+1)}$ (mod $n$) along $\tau=0$ for $x\in X$, while we sew ${\mathbb H}_-^{(j)}$ to 
${\mathbb H}_+^{(j)}$ for $x\notin X$. This gives a conifold ${\cal C}^{(n)}_X$ with a $d-$1-dimensional submanifold of conical singularities along the boundary 
$\partial X\cap\{\tau=0\}$. Then 
$$
P^{(n)}_X=\frac{Z({\cal C}^{(n)}_X)}{Z^n}\,.
$$

In general, for $d>1$, $Z({\cal C}^{(n)}_X)$ has a leading term going like $\exp\!\big(a_n{\rm Vol}(\partial X)\Lambda^{d-1}\big)$ where $\Lambda$ is an ultraviolet cut-off, and the dimensionless coefficients $a_n$ are non-universal. This gives a term in the R\'enyi entropies proportional to $a_n{\rm Vol}(\partial X)\Lambda^{d-1}$ -- the famous `area law' term, which is non-universal and therefore theoretically less interesting. However, in the case where $X$ consists of two disjoint compact regions $(A,B)$ the quantities
\begin{equation}\label{eq:5}
I^{(n)}(A,B)\equiv S^{(n)}_{A}+S^{(n)}_{B}-S^{(n)}_{A\cup B}=(n-1)^{-1}
\log\left(\frac{Z({\cal C}^{(n)}_{A\cup B})Z^n}{Z({\cal C}^{(n)}_A)Z({\cal C}^{(n)}_B)}\right)
\end{equation}
are free of these non-universal contributions, and should depend only on the geometry of $A$ and $B$ and the universal data of the renormalised QFT. 

In general, however computing $Z({\cal C}^{(n)}_{A\cup B})$ is difficult, and, even for $d=1$, we have explicit results only for free field theories \cite{cct,ch}. However in the limit when the linear sizes $R_{A,B}$ of $A$ and $B$ are much smaller than their separation $r$, it was shown in Ref.~\cite{cct} that an expansion in increasing (in general, fractional) powers of $R_AR_B/r^2$ is possible with coefficients which are calculable in principle.  We now generalise this argument to dimensions $d>1$. 

The basic idea is that, from the point of view of an observer far from $A$ or $B$, the sewing together of the copies along the boundaries of $A$ and $B$ should be expressible as a weighted sum of products of local operators $\Phi^{(j)}_{k_j}$ at some conventionally chosen points $(r_A,r_B)$ inside $A$ and $B$, where $\Phi^{(j)}_{k_j}$ is in the algebra of local operators of the QFT defined on $j$th copy of ${\mathbb R}^{d+1}$. That is, the sewing operation in each region can be thought of as a semi-local operator which couples together the $n$ QFTs, but can itself be expressed as a sum of a product of local operators in a direct product of the QFTs.\footnote{The idea of writing correlators of non-local objects in terms of those of local fields has been applied in the past to Wilson loops, see \cite{wilsonloop}.} We write
\begin{equation}\label{eq:2}
\frac{Z({\cal C}^{(n)}_{A\cup B})}{Z^n}=\langle\Sigma^{(n)}_A\Sigma^{(n)}_B\rangle_{({\mathbb R}^{d+1})^n}\,,
\end{equation}
where
\begin{equation}\label{eq:3}
\Sigma^{(n)}_{A}=\frac{Z({\cal C}^{(n)}_{A})}{Z^n}\,\sum_{\{k_j\}}C^{A}_{\{k_j\}}\prod_{j=0}^{n-1}\Phi_{k_j}(r_{A}^{(j)})\,,
\end{equation}
and similarly for $B$.
Here $k_j$ label a complete set of operators on the $j$th copy. The prefactor on the rhs is inserted since we expect the leading behaviour as $r\to\infty$ to come from the term when all the $\Phi_{k_j}$ are the identity operator, and in this limit $\Sigma^{(n)}_{A\cup B}\sim \Sigma^{(n)}_{A}\Sigma^{(n)}_{B}$. 

The main point now is that the coefficients $C^{A,B}_{\{k_j\}}$ should be universal and therefore independent of the other regions and other local operator insertions as long as they are far away. Thus we can compute them in the simpler situation when $X=A$ but we consider the correlation functions of an arbitrary set of local operators at points $r^{(j)}$ on ${\cal C}^{(n)}_A$ outside $A$:
$$
\langle\prod_{j'}\Phi_{k_{j'}'}(r^{(j')})\rangle_{{\cal C}^{(n)}_A}=\left<\left(\prod_{j'}\Phi_{k_{j'}'}(r^{(j')})\right)
\left(\sum_{\{k_j\}}C^{A}_{\{k_j\}}\prod_{j=0}^{n-1}\Phi_{k_j}(r_{A}^{(j)})\right)\right>_{({\mathbb R}^{d+1})^n}\,.
$$
Note that the rhs decomposes into a sum of products of 2-point functions on each copy of ${\mathbb R}^{d+1}$, that is we may take $j'=j$.

This is valid for any QFT, but in the special case of a CFT we can choose the complete set of local operators so that their 2-point functions are orthonormal for \em all \em separations:
$$
\langle\Phi_{k'}(r)\Phi_k(r_A)\rangle=\frac{\delta_{k'k}}{|r-r_A|^{2x_k}}\,,
$$ 
where $x_k$ is the scaling dimension of $\Phi_k$ and we have assumed scalar operators for simplicity.

Thus we find, taking the limit $r^{(j)}\to\infty_j$ (that is, infinity on the the $j$th copy of ${\mathbb R}^{d+1}$)
$$
C^A_{\{k_j\}}=\lim_{\{r^{(j)}\}\to\infty_j}|r^{(j)}|^{\sum_jx_{k_j}}\langle\prod_j\Phi_{k_j}(r^{(j)})\rangle_{{\cal C}^{(n)}_A}\,.
$$
Note that $C^A_{\{k_j\}}\propto R_A^{\sum_jx_{k_j}}$ by dimensional analysis.

In 1+1 dimensions, when $A$ is an interval of finite length, ${\cal C}^{(n)}_A$ may be uniformized to $\mathbb C$ by a conformal mapping and therefore the $C^A_{\{k_j\}}$ are easily computable, at least for the first few leading terms in the expansion. In higher dimensions this is more difficult. Sec.~3 of this paper will be devoted to the simplest case of a free scalar field theory. 

However, supposing that we have computed the $C^A_{\{k_j\}}$, then substituting into (\ref{eq:2},\ref{eq:3}) and again using orthonormality of the operators
$$
\frac{Z({\cal C}^{(n)}_{A\cup B})}{Z^n}=\frac{Z({\cal C}^{(n)}_{A})}{Z^n}\frac{Z({\cal C}^{(n)}_{B})}{Z^n}\sum_{\{k_j\}}
C^A_{\{k_j\}}C^B_{\{k_j\}}\,r^{-2\sum_jx_{k_j}}\,,
$$
so that
\begin{equation}\label{eq:4}
\frac{P^{(n)}_{A\cup B}}{P^{(n)}_{A}P^{(n)}_{B}}=\sum_{\{k_j\}}
C^A_{\{k_j\}}C^B_{\{k_j\}}\,r^{-2\sum_jx_{k_j}}\,,
\end{equation}
where $r=|r_A-r_B|$.  Note that in the ratio all terms which contain the UV divergent area law pieces cancel. 

If we now arrange the operators $\Phi_{k_j}$ in order of increasing dimension $x_{k_j}$, this gives an expansion in increasing powers of $(R_AR_B/r^2)$. The leading term is unity, and comes from taking all the $\Phi_{k_j}$ to be the identity operator $\bf 1$ with $x=0$. In $d=1$, the leading terms with a single $x_{k_j}>0$ in general vanish because the one-point functions of primary operators in ${\cal C}^{(n)}_{A}$ are proportional to those in $\mathbb C$ where they vanish. However this is not necessarily the case for $d>1$, unless the one-point function vanishes for symmetry reasons. The next contribution comes from taking two of the $x_{k_j}>0$. Note that, unlike the case of $d=1$, these do not have to be equal because orthogonality may not hold on ${\cal C}^{(n)}_{A}$. However, the leading correction terms will come from the smallest two non-zero $x_k$, and, barring degeneracies, these will correspond to the same operator. 

As an example consider the 2+1-dimensional Ising field theory. The leading operators are the magnetisation, with $x_\sigma\approx0.52$, and the energy operator with $x_\epsilon\approx1.41$. The one-point function of the magnetisation on ${\cal C}^{(n)}_{A}$ vanishes by the ${\mathbb Z}_2$ symmetry of the model, but there is no reason for the one-point function of the energy operator to vanish. Since, however, $2x_\sigma<x_\epsilon$, the leading term in the mutual information will be proportional to $(R_AR_B/r^2)^{2x_\sigma}$, with a correction of order $(R_AR_B/r^2)^{x_\epsilon}$. 
Although the above inequality holds for many interacting CFTs, counterexamples exist in supersymmetric theories \cite{susy}, in which case the leading term in the will correspond to the one-point function on ${\cal C}^{(n)}_{A,B}$ of an operator with the quantum numbers of the vacuum.

A second example is free scalar field theory in all $d>1$, to be considered in detail in the following section. (For $d=1$ the field itself is not a local operator and the leading corrections then come from exponentials and derivatives of the 
field \cite{cct}.) The field $\phi$ has dimension $x=(d-1)/2$ but has vanishing one-point function, once again because of ${\mathbb Z}_2$ symmetry under $\phi\to-\phi$. However $:\!\phi^2\!:$ has dimension $d-1$ and a non-zero one-point function (see below). The leading term in the mutual information is then a combination of these two contributions, and goes like $(R_AR_B/r^2)^{d-1}$. For a massless Dirac field, which has dimension $d/2$, the power $d-1$ in this expression is replaced by $d$. These results agree with those of Refs.~\cite{ch,shiba}. 

The corrections to this leading behaviour come from larger values of the exponent $\sum_jx_{k_j}$ in (\ref{eq:4}), and from higher terms on the expansion of the logarithm in (\ref{eq:5}). For a free scalar field, the first type of correction comes from when either four of the $\Phi_{k_j}$ are taken to be $\phi$, or when these are taken in pairs as $:\!\!\phi^2\!\!:$. These all give a contribution $O((R_AR_B/r^2)^{2(d-1)})$. Note that these are more important for smaller values of $d$. There is also a correction coming from taking one of the $\Phi_{k_j}$ to be the stress tensor, which always has dimension $(d+1)$, and taking 2 of them to be the current $\partial_\mu\phi$. These both lead to universal corrections $O((R_AR_B/r^2)^{d+1})$. For $d>3$ they dominate the other corrections.

\section{Free scalar field theory}
In this section we consider the case of a free massless scalar field. 
The action is proportional to $\int(\partial\phi)^2d^{d+1}x$, and we normalise the field so that its 2-point function in ${\mathbb R}^{d+1}$ is\footnote{We use $r$ to denote points in ${\mathbb R}^{d}$ and $x=(r,\tau)$ points in ${\mathbb R}^{d+1}$.}
$$
\langle\phi(x_1)\phi(x_2)\rangle\equiv G_0(x_1-x_2)=|x_1-x_2|^{-(d-1)}\,.
$$
As discussed in the previous section, we also need $:\!\phi^2\!:$, which may be defined by point-splitting as 
$$
:\!\phi^2(x)\!:=\lim_{\delta\to0}\left(\phi(x+\ffrac12\delta)\phi(x-\ffrac12\delta)-G_0(\delta)\right)\,.
$$
Its 2-point function in ${\mathbb R}^{d+1}$, by Wick's theorem, is
$$
\langle:\!\phi^2(x_1)\!::\!\phi^2(x_2)\!:\rangle=2G_0(x_1-x_2)^2\,,
$$
so, to conform to our normalisation convention, we should consider $\Phi(x)\equiv2^{-1/2}:\!\phi^2(x)\!:$.

We would like to compute the analogous correlation functions in ${\cal C}^{(n)}_A$. 
As in Ref.~\cite{cct}, rather than thinking of a single free field on this conifold, we think of $n$ copies $\phi_j$ on ${\mathbb R}^{d+1}$, coupled by the boundary conditions across $\tau=0$:
\begin{eqnarray*}
\phi_j(r,0-)&=&\phi_{j+1}(r,0+)\qquad(r\in A)\,;\\
&=&\phi_{j}(r,0+)\qquad(r\notin A)\,.
\end{eqnarray*}
Thus the coefficients $C^A_{\{k_j\}}$ we need to lowest order are 
\begin{equation}\label{eq:7}
C^{A}_{jj'}\equiv C^A_{0,\ldots,1,\ldots,1,\ldots,0}=\lim_{x_1,x_2\to\infty}(x_1x_2)^{d-1}\langle\phi_{j}(x_1)\phi_{j'}(x_2)\rangle_{{\cal C}^{(n)}_A}\,,
\end{equation}
where the non-zero entries occur at $j\not=j'$,
and
\begin{equation}\label{eq:8}
C^{A}_{jj}\equiv C^A_{0,\ldots,2,\dots,0}=2^{-1/2}\lim_{x\to\infty} x^{2(d-1)}\langle:\!\phi_j^2(x)\!:\rangle_{{\cal C}^{(n)}_A}\,.
\end{equation}
In the language of $d$+1-dimensional electrostatics on  ${\cal C}^{(n)}_A$, $\langle\phi_{j}(x_1)\phi_{j'}(x_2)\rangle$ is the potential at
$x_1$ on copy $j$ due to a unit charge at $x_2$ on copy $j'$, while $\langle:\!\!\phi_{j'}^2(x)\!\!:\rangle$ is the excess self-energy of a unit charge at $x$ on copy $j'$. Note that 
$$
\sum_{j\not=j'}\langle\phi_{j}(x_1)\phi_{j'}(x_2)\rangle_{{\cal C}^{(n)}_A}+\langle:\!\!\phi_{j'}^2(x)\!\!:\rangle_{{\cal C}^{(n)}_A}=0\,,
$$
which follows from conservation of electric flux.

The leading correction in (\ref{eq:4}) is then
\begin{eqnarray}
&&r^{-2(d-1)}\big(\ffrac12\sum_{j\not=j'}C^{A}_{jj'}C^{B}_{jj'}+\sum_jC^{A}_{jj}C^{B}_{jj}\big)\nonumber\\
&=&\ffrac12r^{-2(d-1)}\,n\left[\sum_{j=1}^{n-1}\left(\lim_{x_1\to\infty}(x_1^2)^{d-1}\langle\phi_{j}(x_1)\phi_{0}(x_1)
\rangle_{{\cal C}^{(n)}_A}\right)\left(\lim_{x_1\to\infty}(x_1^2)^{d-1}\langle\phi_{j}(x_1)\phi_{0}(x_1)\rangle_{{\cal C}^{(n)}_B}\right)\right.\nonumber\\
&&\qquad\qquad\qquad+\left.
\left(\lim_{x_1\to\infty}(x_1^2)^{d-1}\langle:\!\phi_{0}^2(x_1)\!:
\rangle_{{\cal C}^{(n)}_A}\right)\left(\lim_{x_1\to\infty}(x_1^2)^{d-1}\langle:\!\phi_{0}^2(x_1)\!:\rangle_{{\cal C}^{(n)}_B}\right)
\right]\,,\label{eq:6}
\end{eqnarray}
where we have used the cyclic symmetry to extract an overall factor of $n$.

\subsection{The case $n=2$}

For $n=2$ it is useful to define the linear combinations $\phi_{\pm}=2^{-1/2}(\phi_0\pm\phi_1)$, which satisfy
\begin{eqnarray*}
\phi_-(r,0-)&=&-\phi_-(r,0+)\qquad(r\in A)\,;\\
&=&+\phi_-(r,0+)\qquad(r\notin A)\,,
\end{eqnarray*}
while $\tilde\phi_+$ is continuous across $\tau=0$. Note that
$$
\langle\phi_\pm(x)\phi_\pm(x_1)\rangle=\langle\phi_0(x)\phi_0(x)\rangle\pm\langle\phi_1(x)\phi_0(x)\rangle\,,
$$
so that the correlation functions on the left hand side can be interpreted as the potential at $x$ due to a unit charge at $x_1$. For the upper + sign, the potential is continuous everywhere else, and so is equal to $G_0(x-x_1)$. 

For the lower $-$ sign, however, it is constrained to change sign across $A\cap(\tau=0)$. 
 We notice that in (\ref{eq:7},\ref{eq:8}) we may take $x_1\to\infty$ in any direction. For convenience choose it to lie in the hyperplane $\tau=0$. Then the potential due to a unit charge at $x_1$ must be symmetric under reflection $\tau\to-\tau$. Therefore the potential on $A\cap(\tau=0)$ vanishes. 
 Thus, as far as $\phi_-$ is concerned, $A\cap\{\tau=0\}$ acts like a conductor, at zero electrostatic potential. 

Thus we have the electrostatics problem of finding the potential at $x$ due to a unit charge at $x_1$, in the presence of a conductor held at zero potential at $A\cap\{\tau=0\}$. In general this is complicated, but since we are only interested in the far field in the limit when $|x_1-r_A|\gg R_A$, we can make a simple approximation, valid in this limit. Define  $\bar\phi(x)\equiv\langle\phi_-(x)\phi_-(x_1)\rangle -G_0(x-x_1)$.
Then $\bar\phi(x)$ is regular at $x_1$ and takes an approximately constant value $-|x_1|^{-(d-1)}$ on the conductor. This will induce a total charge $-{\bf C}_A|x_1|^{d-1}$ on the conductor, where ${\bf C}_A$ is its electrostatic capacitance. Therefore, as $x,x_1\to\infty$,
$\langle\phi_-(x)\phi_-(x_1)\rangle -G_0(x-x_1)\sim-{\bf C}_A|x|^{-(d-1)}|x_1|^{-(d-1)}$. Thus
\begin{eqnarray}
\langle\phi_1(x_1)\phi_0(x_1)\rangle&\sim&\ffrac12{\bf C}_A|x_1|^{-2(d-1)}\,;\label{14}\\
\langle:\!\phi_0^2(x_1)\!:\rangle&=&\lim_{x\to x_1}\big(\langle\phi_0(x)\phi_0(x_1)\rangle-G_0(x-x_1)\big)=
-\ffrac12{\bf C}_A|x_1|^{-2(d-1)}\,,\label{15}
\end{eqnarray}
giving
\begin{equation}\label{11}
I^{(2)}(A,B)\sim \frac{{\bf C}_A{\bf C}_B}{2r^{2(d-1)}}\,.
\end{equation}

This is valid for any compact regions $A$ and $B$. On dimensional grounds ${\bf C}_{A,B}\propto R_{A,B}^{d-1}$, but the coefficient depends on the shape of the regions. Very few cases are known exactly. 

In the Appendix we show, generalising a result of W.~Thomson (Lord Kelvin), that when $\partial A$ is a hypersphere $S^{d-1}$of radius $R_A$ , so that $A\cap\{\tau=0\}$ is a disc,
\begin{equation}\label{12}
{\bf C}_A=\frac{\Gamma(d/2)\Gamma(1/2)}{\pi\Gamma((d+1)/2)}\,R_A^{d-1}\,.
\end{equation}
For $d+1=3$ this gives Thomson's result $\frac2\pi R_A$, while for $d+1=4$ it gives $\frac12R_A^2$.

\subsection{Free field theory when $A$ and $B$ are spherical, general $n$.}
The above symmetry argument does not seem to generalise to larger values of $n$, but further analytic progress can be made in the case when $A$ and $B$ are the interior of spheres $S^{d-1}$. In that case we can exploit the conformal invariance of the free field theory to compute the coefficients $C_{jj'}^{A,B}$. 

Before doing this, we note that conformal invariance implies in this case that $I^{(n)}(A,B)$ is a universal function of the quantity
$$
\frac{R_AR_B}{r^2-(R_A-R_B)^2}\,.
$$
Given any two spheres $S_{d-1}$: $A$ and $B$, we may expand them into spheres $S_d$ of the same radii in $d+1$ dimensions about their common equatorial plane $\tau=0$. Conformal transformations in $d+1$ dimensions will transform them into other spheres (counting hyperplanes as spheres through the point at infinity.) Since the system has axial symmetry about the line joining their centres, we may restrict to conformal transformations which preserve this line. Under such transformations, the cross-ratio of the points $(-\frac12r-R_A,-\frac12r+R_A,\frac12r-R_B,
\frac12r+R_B)$ where the spheres intersect this line is invariant. This is the quantity above, apart from a factor of 4
. Since by (\ref{eq:5}) 
$I^{(n)}(A,B)$ is given in terms of a ratio of partition functions in which all metrical factors cancel, it should be both scale and conformally invariant.

We now use conformal invariance to compute correlation functions on ${\cal C}^{(n)}_A$ when $A$ is a sphere.
We may regard $A$ as the intersection of a $d$-dimensional ball of radius $R_A$ with the equatorial plane $\tau=0$. Consider the effect of making an inversion in ${\mathbb R}^{d+1}$ which sends a point on the boundary of $A$ to the point at infinity. This maps the boundary of the ball $S^d$ into a hyperplane ${\mathbb R}^d$ . The plane $\tau=0$ is preserved by the mapping, and so the boundary of ${\cal C}_A$ is mapped into a hyperplane ${\mathbb R}^{d-1}$. $A\cap\{\tau=0\}$ itself is mapped into a $d$-dimensional half-space. This is easier to visualise in $d=2$, when $A\cap\{\tau=0\}$ is mapped into a half-plane with an infinite line ${\mathbb R}$ as its boundary. In the replicated theory this turns into a line of conical singularities.  The conifold ${\cal C}_A^{(n)}$ is mapped into ${\cal C'}_A^{(n)}=\{\mbox{2-dimensional cone of opening angle $2\pi n$}\}\times{\mathbb R}^{d-1}$.

The correlation functions transform covariantly under this conformal mapping:
$$
\langle\phi_j(x_1)\phi_{j'}(x_2)\rangle_{{\cal C}_A^{(n)}}=\left|\frac{\partial x_1'}{\partial x_1}\right|^{(d-1)/2}
\left|\frac{\partial x_2'}{\partial x_2}\right|^{(d-1)/2}\,\langle\phi_j(x'_1)\phi_{j'}(x'_2)\rangle_{{\cal C'}_A^{(n)}}\,.
$$
The mapping brings the points at $x_{1,2}=\infty$ to a finite distance $1/(2R_A)$ from the conical singularity. The Jacobian cancels the factors of $(x_1x_2)^{d-1}$ in (\ref{eq:7},\ref{eq:8}). Thus 
$$
C^A_{jj'}=\langle\phi_j(1/2R_A)\phi_{j'}(1/2R_A)\rangle_{{\cal C'}_A^{(n)}}=(2R_A)^{d-1}\langle\phi_j(1)\phi_{j'}(1)\rangle_{{\cal C'}_A^{(n)}}\,,
$$
and similarly for $C^A_{jj}$.

Thus we need to compute the potential on the $j$th copy at unit distance from the hyperplane of conical singularities due to a unit charge in the same position on the $j'$th copy. As before, because of the cyclic symmetry we can take $j'=0$. Since this problem now has axial symmetry the calculation is simplified. One approach is to introduce cylindrical polar coordinates $(\rho,\theta,\vec z)$, where $\theta\in[0,2\pi n]$ and $\vec z$ is a $(d-1)$-dimensional coordinate in the ${\mathbb R}^{d-1}$ subspace. We need the Green's function satisfying
$$
-\nabla^2G^{(n)}(\rho,\theta,\vec z)\propto\delta(\rho-1)\delta(\theta)\delta^{d-1}(\vec z)\,,
$$
with $G^{(n)}(\rho,\theta+2\pi n,\vec z)=G^{(n)}(\rho,\theta,\vec z)$. We then have 
$\langle\phi_j(1)\phi_{0}(1)\rangle_{{\cal C'}_A^{(n)}}=G^{(n)}(1,2\pi j/n,0)$. 
An expression for $G(\rho,\theta,\vec z)$ may be found by Fourier transforming with respect to $\vec z$ and solving in terms of Bessel functions, but the resultant integrals and sums are ill-conditioned and we have not been able to make the continuation in $n$.

We adopt a different approach, for which the complete answer for all $n$ may be obtained for all $d$.
Instead of considering $n$ to be initially a positive integer, suppose $n=1/m$ where $m$ is a positive integer. The solution for
$0\leq\theta\leq2\pi/m$ is then immediate  by the method of images:
$$
G^{(1/m)}(\rho,\theta,z)=\sum_{k=0}^{m-1}G_0(\rho,\theta+2\pi k/m,z)\,.
$$
Specialising to $\rho=1$, $z=0$,
\begin{equation}\label{eq:9}
G^{(1/m)}(1,\theta,0)=\sum_{k=0}^{m-1}\frac1
{\big(2-2\cos(\theta+2\pi k/m)\big)^{(d-1)/2}}\,.
\end{equation}
For $d-1$ even this sum is straightforward, but more difficult for the odd case, as we illustrate below.

\subsubsection{The case $d=3$}
In this case the sum can be evaluated explicitly in a number of ways. For example, we can regularise it and write it as 
$$
\lim_{\rho\to1-}\sum_{k=0}^{m-1}\frac1{\big(1-\rho e^{i(\theta+2\pi k/m)}\big)\big(1-\rho e^{-i(\theta+2\pi k/m)}\big)}
=\lim_{\rho\to1-}\sum_{k=0}^{m-1}\sum_{p=0}^\infty\sum_{p'=0}^\infty\rho^{p+p'}e^{i(p-p')(\theta+2\pi k/m)}\,,
$$
where $|\rho|<1$.
The sum over $k$ vanishes unless $p-p'=0$ (mod $m$), when it gives $m$. For $p\geq p'$ we can write $p=p'+lm$, and similarly for $p'\geq p$. Subtracting off the $l=0$ term to avoid double counting gives
$$
m\sum_{p'=0}^\infty\sum_{l=0}^\infty\rho^{2p'+lm}e^{ilm\theta}+{\rm c.c.}-m\sum_{p'=0}^\infty\rho^{2p'}
=\frac m{1-\rho^2}\left(\frac1{1-\rho^m e^{im\theta}}+\frac1{1-\rho^m e^{-im\theta}}-1\right)\,.
$$
Taking the limit $\rho\to1-$ then gives the simple result for $d=3$
$$
G^{(1/m)}(1,\theta,0)=\frac{m^2}{2-2\cos m\theta}\,.
$$
This valid for $m$ a positive integer. However, since it vanishes exponentially fast as $m\to\pm i\infty$, Carlson's theorem
ensures that it has a well-defined analytic continuation, to other values of $m$, in particular to $m=1/n$.  Note however that this does not make sense before performing the sum in (\ref{eq:9})!

First, we note that
\begin{equation}\label{eq:phi2}
\langle:\!\phi_0^2(1)\!:\rangle_{{\cal C'}_A^{(n)}}=\lim_{\theta\to0}\left(\frac{1/n^2}{2-2\cos(\theta/n)}-
\frac{1}{2-2\cos\theta}\right)=\frac{1-n^2}{12n^2}\,.
\end{equation}
As we show in Sec.~3.3, this relates to the coefficient of a universal term in the R\'enyi entropy of a single sphere in the massive theory. Note that this does not contribute to the mutual information through (\ref{eq:6}), since this term would be $O((n-1)^2)$ and therefore have vanishing derivative at $n=1$. A similar remark applies to the 1-point function of the stress tensor $\langle T\rangle_{{\cal C}_A^{(n)}}$.

The first term in (\ref{eq:6}) involves the sum
$$
\sum_{j=1}^{n-1}G^{(n)}(1,2\pi j/n,0)^2=\frac1{n^4}\sum_{j=1}^{n-1}\frac1{\big(2-2\cos(2\pi j/n)\big)^2}\,.
$$
This may be evaluated for $n$ a positive integer by first regulating it as above, and expanding in powers of $\rho$:
$$
\sum_{j=1}^{n-1}\sum_{p=0}^\infty\sum_{p'=0}^\infty(p+1)(p'+1)\rho^{p+p'}e^{2\pi i(p-p')j/n}\,.
$$
The sum over $j$ now gives $-1$ unless $p-p'=0$ (mod $n$), when it gives $n-1$. In this case, writing $p=p'+ln$, etc., as  before,
we get
\begin{eqnarray*}
2n\sum_{p=0}^\infty\sum_{l=0}^\infty(p+1)(p+1+ln)\rho^{2p+ln}-n\sum_{p=0}^\infty(p+1)^2\rho^{2p}-\sum_{p,p'=0}^\infty
(p+1)(p'+1)\rho^{p+p'}\\
=\left(\frac{2n}{(1-\rho^2)^3}-\frac n{(1-\rho^2)^2}\right)\frac{1+\rho^n}{1-\rho^n}+\frac{2n^2\rho^n}{(1-\rho^2)^2(1-\rho^n)^2}
-\frac1{(1-\rho)^4}\,.
\end{eqnarray*}
Taking the limit $\rho\to1$ gives, after some algebra,
$$
\sum_{j=1}^{n-1}G^{(n)}(1,2\pi j/n,0)^2=\frac{(n^2-1)(n^2+11)}{720n^4}\,.
$$
Once again Carlson's theorem assures us of a unique continuation to non-integer values of $n$.

Putting all the pieces together we find, for $d=3$
\begin{eqnarray*}
I^{(n)}(A,B)&\sim&\frac{n}{2(n-1)}r^{-4}\left(\frac{(n^2-1)(n^2+11)}{720n^4}+\frac{(n^2-1)^2}{144n^4}\right)(2R_A)^2
(2R_B)^2\\
&=&\frac{n^4-1}{15n^3(n-1)}\left(\frac{R_AR_B}{r^2}\right)^2\,.
\end{eqnarray*}
Note that for $n=2$ this agrees with (\ref{11}), using (\ref{12}). This is a non-trivial check of our methods.
Taking the derivative at $n=1$ we find for the mutual information the leading term
$$
I(A,B)\sim\frac4{15}\left(\frac{R_AR_B}{r^2}\right)^2\,.
$$

Clearly a similar calculation can be done for any even $d-1$. The result will always be a rational function of $\rho$ and $\rho^n$, which will give, on taking the limit $\rho\to1$, a polynomial in $n$.

\subsubsection{The case $d=2$}
In this case the sum in (\ref{eq:9}) is
$$
G^{(1/m)}(1,\theta,0)=\sum_{k=0}^{m-1}\frac1{\big(2-2\cos(\theta+2\pi k/m)\big)^{1/2}}=
\sum_{k=0}^{m-1}\frac1{2\sin(\frac\theta 2+\frac{\pi k}m)}\,.
$$
(Note that in the physical region $0<\theta<2\pi/m$ the sine is always positive.) Now use the integral representation
$$
\frac1{\sin\pi\mu}=\frac1\pi\int_0^\infty\frac{x^{\mu-1}}{1+x}dx\qquad(0<\mu<1)\,.
$$
Inserting this and performing the sum gives
\begin{equation}\label{13}
G^{(1/m)}(1,\theta,0)=\frac1{2\pi}\int_0^\infty\frac{x^{(\theta/2\pi)-1}(1-x)}{(1+x)(1-x^{1/m})}dx\,.
\end{equation}
The analyticity properties of this expression in $1/m$ may be inferred by dividing the integration region into $(0,1)$ and $(1,\infty)$ and expanding in powers of $x$ ($1/x$) in each case. This gives
$$
\sum_{p=0}^\infty\sum_{q=0}^\infty(-1)^p\left[\frac1{\big(\frac\theta{2\pi}+p+\frac qm\big)\big(\frac\theta{2\pi}+p+\frac qm+1\big)}
+\frac1{\big(-\frac\theta{2\pi}+p+\frac{q+1}m\big)\big(-\frac\theta{2\pi}+p+\frac{q+1}m+1\big)}\right]\,.
$$
In the physical region for $\theta$ this has an accumulation of poles for negative real $m$, but is otherwise analytic. Moreover it grows like $|m|$ as $m\to\infty$ except along the negative real axis. We infer from this that it has a unique continuation to the whole $m$ plane apart from the negative real axis, in particular to $m=1/n$ where $n\geq 1$, which is found by simply setting $m=1/n$ in (\ref{13}). We remark in passing that as $|n|\to\infty$ the sum is $O(n^{-2}\log n)$, consistent with there being a branch cut along the negative real axis. 

 We then see that
$$
\langle:\!\phi_0(1)^2\!:\rangle_{{\cal C'}_A^{(n)}}=\lim_{\theta\to0}\frac1{2\pi}\int_0^\infty\frac{x^{(\theta/2\pi)-1}}{1+x}\left(
\frac{1-x}{1-x^n}-1\right)dx
=-\frac1{2\pi}\int_0^\infty\frac{1-x^{n-1}}{(1+x)(1-x^n)}dx\,.
$$
By substituting $x\to x^{-1}$, we see that integral is twice its value with an upper limit of 1. For $n=2$ we get $-(1/2\pi)$, consistent with (\ref{14},\ref{15},\ref{12}). As $n\to1$
$$
\langle:\!\phi_0(1)^2\!:\rangle_{{\cal C'}_A^{(n)}}\sim\frac{(n-1)}\pi\int_0^1\frac{\log x}{1-x^2}
=-\frac{(n-1)}\pi\sum_{p=0}^\infty\frac1{(2p+1)^2}=-\frac\pi 8(n-1)\,.
$$

The sum in the first term in (\ref{eq:6}) is
\begin{eqnarray*}
\sum_{j=1}^{n-1}G^{(n)}(1,2\pi j,0)^2=\frac1{4\pi^2}\sum_{j=1}^{n-1}\int_0^\infty\int_0^\infty\frac{(xy)^{j-1}(1-x)(1-y)}
{(1+x)(1+y)(1-x^n)(1-y^n)}dxdy\\
=\frac1{4\pi^2}\int_0^\infty\int_0^\infty\frac{(1-x)(1-y)(1-(xy)^{n-1})}
{(1+x)(1+y)(1-xy)(1-x^n)(1-y^n)}dxdy\,.
\end{eqnarray*}
For $n=2$ this is $1/4\pi^2$, consistent with (\ref{14},\ref{15},\ref{12}). As $n\to1$ we find
$$
-\frac{n-1}{4\pi^2}\int_0^\infty\int_0^\infty\frac{\log(xy)}{(1+x)(1+y)(1-xy)}dxdy\,.
$$
This integral may be done by letting $y=u/x$ and first carrying out the $x$-integral, yielding
$$
\frac{n-1}{4\pi^2}\int_0^\infty\frac{(\log u)^2}{(1-u)^2}du=\ffrac16(n-1)\,.
$$
Putting together the pieces, we then find for the leading term in the mutual information for $d=2$
$$
I(A,B)\sim\ffrac12\cdot\ffrac16\,r^{-2}(2R_A)(2R_B)=\frac13\left(\frac{R_AR_B}{r^2}\right)\,.
$$
\subsection{Universal correction for a single spherical region}
Although the main purpose of this section is to study the mutual information of two spherical regions, it is worth noting that the result in (\ref{eq:phi2}) may be used to find a universal logarithmic correction to the area law for a single spherical region in $d=3$. 

Suppose $A$ is the interior of a sphere $S^{d-1}$ of radius $R$ centred at the origin. For the time being we keep $d$ general. Let $\vec R=(R,0,\ldots)$ in the plane $\tau=0$, and invert $x=(r_1,\vec r_\perp,\tau)$ in $d+1$ dimensions, sending the point
$-\vec R$ to infinity:
\begin{eqnarray*}
r_1'&=&\frac{r_1+R}{(r_1+R)^2+\vec r_\perp^2+\tau^2}-\frac1{2R}\,,\\
\vec r'_\perp&=&\frac{\vec r_\perp}{(r_1+R)^2+\vec r_\perp^2+\tau^2}\,.\\
\tau'&=&\frac{\tau}{(r_1+R)^2+\vec r_\perp^2+\tau^2}\,.
\end{eqnarray*}
As before, this maps the conifold ${\cal C}^{(n)}_A$ into ${\cal C'}^{(n)}_A=\{\mbox{cone in $(r_1',\tau')$}\}\times{\mathbb R}_\perp^{d-1}$, and
$$
\langle:\!\phi^2(x)\!:\rangle_{{\cal C}^{(n)}_A}=\left|\frac{\partial x'}{\partial x}\right|^{d-1}\,\langle:\!\phi^2( x')\!:\rangle_{{\cal C'}^{(n)}_A}\,.
$$
The jacobian pre-factor is
$$
(x^2)^{-(d-1)}= [(r_1+R)^2+\vec r_\perp^2+\tau^2]^{-(d-1)}\,,
$$
and 
$$
\langle:\!\phi^2_j(x')\!:\rangle_{{\cal C'}^{(n)}_A}=a_n({r'_1}^2+{\tau'}^2)^{-(d-1)/2}
$$
where, from (\ref{eq:phi2}), 
$$
a_n=\langle:\!\phi^2_j(1)\!:\rangle_{{\cal C'}^{(n)}_A}=(1-n^2)/(12n^2)\,.
$$

After some algebra we find
\begin{equation}\label{eq:phi2Cn}
\langle:\!\phi^2_j(x)\!:\rangle_{{\cal C}^{(n)}_A}
=a_n\left(\frac{4R^2}{[r^2+\tau^2-R^2]^2+4R^2\tau^2}\right)^{(d-1)/2}\,.
\end{equation}
Note that the final answer depends only on $r^2=x_1^2+x_\perp^2$ and $\tau$, as it should. At large $|x|$ it decays as
$(x^2)^{-(d-1)}$ so is integrable at infinity if $d>3$. On the other hand, close to the conical singularity $(r\sim R, \tau\sim0)$, writing $r=R+u$ with
$|u|\ll R$, it behaves as $(u^2+\tau^2)^{-(d-1)/2}$, so is integrable if $d<3$. 

Therefore for $d=3$ the integral of $\langle:\!\phi^2_j(x)\!:\rangle_{{\cal C}^{(n)}_A}
$ over the conifold is logarithmically divergent in the massless theory both in the UV and the IR. The former divergence necessitates a short-distance cut-off $\Lambda^{-1}$. The IR divergence may be regulated by assuming the theory has a small mass $m\ll R^{-1}$. However, since both divergences are logarithmic, to leading order we may still use the expression
(\ref{eq:phi2Cn}). This gives from large distances
$$
a_n\int^{1/m}\frac{4R^2}{x^4}d^4x\sim a_n\cdot(4R^2)(2\pi^2)\log(1/m)\,,
$$
and from short distances
$$
a_n(4\pi R^2)\int_{1/\Lambda}\frac{dud\tau}{u^2+\tau^2}\sim a_n(4\pi R^2)(2\pi)\log\Lambda\,.
$$
Happily the two coefficients agree, so we get
$$
\int\langle:\!\phi^2(\vec x)\!:\rangle_{{\cal C}^{(n)}_A}d^3rd\tau\sim na_n(8\pi^2R^2)\log(\Lambda/m)\,.
$$
The extra factor of $n$ comes from summing over $j$.

The action for the massive theory is
$$
S=\frac1{8\pi^2}\int\big((\partial\phi)^2+m^2\phi^2\big)d^4x\,,
$$
(note the factor of $(4\pi^2)^{-1}$ inserted to conform with our field normalisation convention so that the propagator is
$1/x^2$ when $m=0$.)

Then
$$
(\partial/\partial m^2)\log\big(Z({\cal C}^{(n)}_A)/Z^n\big)=-(8\pi^2)^{-1}\int\langle:\!\phi^2(x)\!:\rangle_{{\cal C}^{(n)}_A}
d^4x\,,
$$
which gives a universal logarithmic correction term to the area law for $mR\ll1$:
$$
\Delta S_A^{(n)}=-\frac{n+1}{12n} R^2m^2\log(\Lambda/m)\,.
$$

Since the dependence on $\Lambda$ comes from the region near the conical singularity, we may conjecture that this is a special case of 
$$
\Delta S_A^{(n)}=-\frac{n+1}{12n}\,\frac{\mbox{Area}(\partial A)}{4\pi}\,m^2\log(\Lambda/m)\,,
$$
valid for any region $A$ with a smooth boundary $\partial A$. This agrees with a calculation in \cite{cc} for the case when $A$ is a half-space (see also \cite{ch})
based on the 1+1-dimensional result:
$$
S_A^{(n)}\sim\frac{n+1}{12n}\mbox{Area}(\partial A)\int\log(k^2+m^2)^{1/2}\,\frac{d^2k}{(2\pi)^2}\,,
$$
after differentiation with respect to $m^2$.

We emphasise that this universal logarithmic term is present only in the massive theory, and is not the same as the term
$O(\log(\Lambda R))$  discussed in \cite{logs} (see also \cite{ch}). This can be obtained using our approach by computing the 1-point function of the stress tensor on the conifold ${\cal C}_A^{(n)}$ \cite{JCinprep}. Although we have carried out our calculation for $d=3$, it may be generalised to any even $d+1$. For $d+1$ odd, logarithms of the type we have discussed are absent.

We also point out that in lattice regularisations we also expect `unusual' corrections coming from (possibly relevant) operators, which do not break the symmetry, localised on the conical singularity \cite{unusual}. In 1+1 dimensions these have scaling dimensions $(x/n)$ and lead to corrections $O(r^{-2x/n})$ in the mutual R\'enyi entropies, where $x$ is the usual bulk scaling dimension of these operators.
For a general CFT in higher dimensions, the $n$-dependence is more complicated. However for a free field theory the 1+1- dimensional result should still hold, and the $:\!\phi^2\!:$ operator on the conical singularity has dimension $(d-1)/n$.
This will lead to corrections $O(r^{-2(d-1)/n})$ in the mutual R\'enyi entropies, which, if present, actually dominate the universal terms we have found for $n>1$. However the amplitude is $O((n-1)^2)$ and therefore they do not contribute to the mutual information.

\section{Conclusions}
We have presented analytic calculations of the mutual R\'enyi information $I^{(n)}(A,B)$ of two disjoint regions in the ground state of a conformal field theory in general space dimension $d$. They are given as an expansion in powers of the  ratio 
$R_AR_B/r^2$ with powers which are integer combinations of the scaling dimensions of the local operators in the theory, with universal coefficients, which factorise between $A$ and $B$. 

Unlike the case of $d=1$ we are able to obtain explicit results only for a free scalar field theory, although similar methods should work for other free theories, and even interacting theories using the $\epsilon$-expansion. For $n=2$ we showed that the coefficient of the leading term is proportional to the product of the capacitances of $A$ and $B$ in $d+1$-dimensional electrostatics. For spheres we computed this by generalising an argument of W.~Thomson to general $d$. 

When $A$ and $B$ are spheres we can obtain explicit results for the coefficient of the leading term for all $n$ and $d$,
although the computations increase in complexity with $d$. For $d+1$ even the results are polynomials in $n$, while for $d+1$ odd they may be expressed as Dirichlet series. We remark that from a technical point of view it was simpler first to  compute the R\'enyi entropies for $n=1/m$ where $m$ is a positive integer, and then perform the analytic continuation. 
This approach may be more generally useful. 

It would be interesting to compare our results with those predicted by the holographic approach of Ryu and Takayangi \cite{rt}, which were hypothesised to extend to the case of more than one region in \cite{hr} (however, see \cite{ton}). We note that our result for $n=2$ that the mutual R\'enyi information is given by the product of the capacitances suggests that it cannot be interpreted as an extensive quantity. 

Although we have considered only the massless case for the mutual information, our results extend straightforwardly to QFTs with a mass scale $m$, as long as $R_A,R_B\ll m^{-1}$, by replacing massless propagator $r^{-(d-1)}$ by its massive version. We also confirmed the existence of a universal correction $O(R^2m^2\log m)$ for the R\'enyi entropies of a single spherical region in the massive theory with $mR\ll1$.

\em Acknowledgements. \em I would like to thank Noburo Shiba for sending me an early version of his paper and for further correspondence. I thank Pasquale Calabrese for his comments on a draft of the present paper, and Mark Srednicki and Erik Tonni for discussions. I also thank Slava Rychkov for drawing my attention to the result in Ref.~\cite{susy}.

\appendix
\section{Thomson's argument for general $d$}
Consider electrostatics in $D=d+1$ dimensions. What is the capacitance of a hollow ellipsoidal shell with semi-axes $a_1,\ldots,a_D$? In the limit $a_D\to0$ this will give the formula for a flat disc in the shape of a $d$-dimensional ellipsoid.
We need to know what charge distribution produces zero field inside. We generalise an argument due to W.~Thomson \cite{thom} (see \cite{pear} for a modern account.)

\begin{figure}[hh]
\centering
\includegraphics[width=0.3\textwidth]{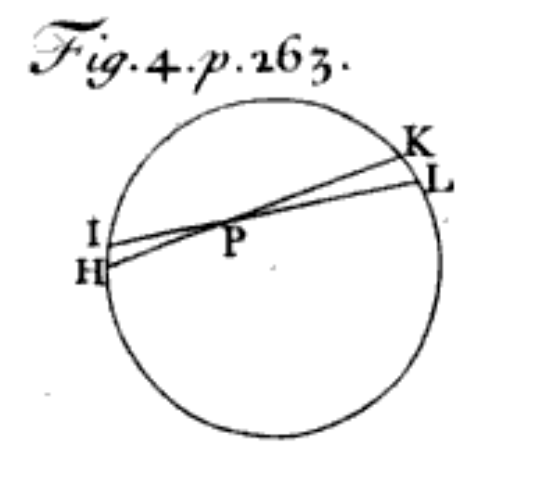}
\caption{\label{fig:newt}Newton's demonstration that the field at any point in the interior of a spherically symmetric charge distribution vanishes. [Reproduced from \em Newton's Principia\em, www.gutenberg.org/files/28233].}
\end{figure}

Start from Newton's observation \cite{newt} that for a sphere, when all the $a_j$ are equal, the field vanishes when the charge distribution is uniform\footnote{Of course Newton was considering gravity, but the argument is valid for any $r^{-d}$ force law in $d+1$ dimensions. }. This may be understood as follows: consider a point $P$ inside the sphere, and a bi-cone of small solid angle $d\Omega$ which intersects the sphere in two regions of areas proportional to $|IH|^d$ and $|KL|^d$ shown in Fig.~\ref{fig:newt}. The contribution to the electric field at $P$ from these two regions
is proportional to
$$
\frac{|IH|^d}{|{PH}|^d}-\frac{|KL|^d}{|PK|^d}
$$
directed along $HP$. But this vanishes by similarity of the two triangles. 

Note that this extends to the case of any spherically symmetric charge distribution. In particular, we can consider a uniform charge distribution between two shells of radii $a$ and $a+\delta a$.

Now consider the effect of making a uniform shear transformation $x_j\to x_j'=\lambda_jx_j$ where $\prod_j\lambda_j=1$. This distorts the two spheres into similar ellipsoids with axes $\lambda_ja$ and $\lambda_j(a+\delta a)$. Gauss' law remains true as long as we rescale the electric field components $E_j\to\lambda_jE_j$, so the electric field inside the ellipsoid still vanishes. The bulk charge density between the ellipsoids remains uniform.

Now take the limit
$\delta a\to0$. We need to work out the thickness of the shell at a point $\{x_j'\}$ on the inner shell, which will be proportional to the surface charge density. This came from a point $\{x_j'/\lambda_j\}$ on the sphere, and hence its image on the outer shell is $x_j'(1+\delta a/a)$. To get the thickness we need to form the inner product of $x_j(\delta a/a)$ with the normal at the point $x_j'$ . The equation of the ellipsoid is
$$
\sum_j(x_j^2/\lambda_j^2)=a^2\,,
$$
so the normal vector satisfies $\sum_jn_jdx_j=0$ where $dx_j$ is any vector satisfying $\sum_jx_jdx_j/\lambda_j^2=0$. The solution is to take $n_j\propto x_j/\lambda_j^2$, and so the thickness at $x_j$ is proportional to 
$$
x_jn_j=\frac{\sum_j(x_j^2/\lambda_j^2)}{\big(\sum_j(x_j^2/\lambda_j^4)\big)^{1/2}}\propto\big(\sum_j(x_j^2/\lambda_j^4)\big)^{-1/2}\,.
$$
This is the surface charge density required to ensure that the field within an ellipsoidal shell vanishes. 

In our case we need to take the limit $\lambda_D\to\infty$. Rescale $x_D=b\lambda_D$ where $|b|<1$. Then the ellipsoid degenerates into 
$$
\sum_{j=1}^d\frac{x_j^2}{a_j^2}=1-b^2<1\,,
$$
and the charge density is
$$
\propto \frac{\lambda_D^2}{x_D}\propto \frac1{|b|}=\frac1{\sqrt{1-\sum_{j=1}^d\frac{x_j^2}{a_j^2}}}\,.
$$

For the case of a spherical disc when all the $a_j=a$ we get $\sigma\propto1/\sqrt{a^2-r^2}$ for any $d$. In this case the total  charge is 
$$
Q\propto\int_0^a\frac{r^{d-1}}{\sqrt{a^2-r^2}}dr=a^{d-1}\int_0^{\pi/2}(\sin\theta)^{d-1}d\theta=a^{d-1}\frac{\Gamma(d/2)\Gamma(1/2)}{2\Gamma((d+1)/2)}\,.
$$

On the other hand the potential is (with the same constant of proportionality)
$$
V=\int_0^a\frac{r^{d-1}}{r^{d-1}\sqrt{a^2-r^2}}dr=\int_0^{\pi/2}d\theta=(\pi/2)\,,
$$
so 
$$
Q=\frac{\Gamma(d/2)\Gamma(1/2)}{\pi\Gamma((d+1)/2)}a^{d-1}\,V={\bf C}V\,.
$$
For $D=3$ this reduces to Thomson's result ${\bf C}=(2/\pi)a$.

\end{document}